\newif\ifpnas
\pnastrue
\pnasfalse

\ifpnas
\documentclass[10pt,oneside,9pt,twocolumn,twoside,lineno]{pnas-new}
\else
\documentclass[10pt,twocolumn,superscriptaddress]{revtex4-2}
\fi
\usepackage{fancyhdr}
\usepackage{color}
\usepackage{units}
\usepackage{graphicx}
\usepackage{amsmath}
\usepackage{amsthm}
\usepackage{amssymb}
\usepackage{enumitem}
\usepackage{hyperref}

\ifpnas

\title{Field-tuned ferroquadrupolar quantum phase transition in the insulator TmVO$_{4}$}

\author[a,1]{Pierre Massat}
\author[b]{Jiajia Wen}
\author[a,b]{Jack M. Jiang}
\author[c]{Alexander T. Hristov}
\author[d,e]{Yaohua Liu}
\author[b,f]{Rebecca W. Smaha}
\author[g]{Robert S. Feigelson}
\author[a,b]{Young S. Lee}
\author[h]{Rafael M. Fernandes}
\author[a,2]{Ian R. Fisher}

\affil[a]{Department of Applied Physics, Stanford University, Stanford, CA
94305, USA}
\affil[b]{Stanford Institute for Materials and Energy Sciences, SLAC National Accelerator Laboratory, 2575 Sand Hill Road, Menlo Park, CA 94025, USA}
\affil[c]{Department of Physics, Stanford University, Stanford, CA 94305}
\affil[d]{Neutron Scattering Division, Oak Ridge National Laboratory, Oak Ridge, TN 37831, USA}
\affil[e]{Second Target Station, Oak Ridge National Laboratory, Oak Ridge, Tennessee 37831, USA}
\affil[f]{Department of Chemistry, Stanford University, Stanford, CA 94305, USA}
\affil[g]{Department of Materials Science and Engineering, Stanford University, Stanford, CA 94305, USA}
\affil[h]{School of Physics and Astronomy, University of Minnesota, Minneapolis,
MN 55455, USA}

\significancestatement{The behavior of metals close to an electronic nematic quantum critical
point (QCP) is not well understood. Here we study the simpler case
of an insulator that undergoes a ferroquadrupolar phase transition,
which is a specific realization of Ising-nematic order. The ferroquadrupolar
transition can be tuned continuously towards a QCP via application
of a magnetic field. We find that close to the QCP the phase boundary
deviates from semi-classical predictions due to quantum fluctuations.
The observed power-law dependence is different from that of magnetic
Ising systems due to coupling of order parameter fluctuations to the
crystal lattice fluctuations (phonons). These results provide a well-understood starting
point to approach more complex cases of nematic quantum criticality
in metallic and disordered materials.}

\leadauthor{Massat}

\authorcontributions{P. M. and R. S. F. grew the samples; P. M., A. T. H., J. W.,
J. M. J., Y. L., R. W. S., Y. S. L., and I. R. F. designed the experiments; P. M. and
A. T. H. carried out the heat capacity and magneto-caloric effect
experiments; J. W., J. M. J., Y. L. and P. M. carried out the neutron experiment;
R. W. S. carried out the single-crystal X-Ray diffraction experiment;
R. M. F. derived the theoretical results; all of the authors contributed
to writing the manuscript.}

\authordeclaration{The authors declare no competing interests.}

\correspondingauthor{E-mails: \textsuperscript{1}pmassat@stanford.edu \textsuperscript{2}irfisher@stanford.edu}


\templatetype{pnasresearcharticle} 

\dates{This manuscript was compiled on \today}
\doi{\url{www.pnas.org/cgi/doi/10.1073/pnas.XXXXXXXXXX}}

\else
\fi

\DeclareMathOperator{\atanh}{atanh}

\begin{document}

\ifpnas
\maketitle

\else

\title{Field-tuned ferroquadrupolar quantum phase transition in the insulator TmVO$_{4}$}
\date{\today}

\author{Pierre Massat}
\email{E-mail: pmassat@stanford.edu}
\affiliation{Department of Applied Physics, Stanford University, Stanford, CA 94305, USA}

\author{Jiajia Wen}
\affiliation{Stanford Institute for Materials and Energy Sciences, SLAC National Accelerator Laboratory, 2575 Sand Hill Road, Menlo Park, CA 94025, USA}

\author{Jack M. Jiang}
\affiliation{Department of Applied Physics, Stanford University, Stanford, CA 94305, USA}
\affiliation{Stanford Institute for Materials and Energy Sciences, SLAC National Accelerator Laboratory, 2575 Sand Hill Road, Menlo Park, CA 94025, USA}

\author{Alexander T. Hristov}
\affiliation{Department of Physics, Stanford University, Stanford, CA 94305}

\author{Yaohua Liu}
\affiliation{Neutron Scattering Division, Oak Ridge National Laboratory, Oak Ridge, TN 37831, USA}
\affiliation{Second Target Station, Oak Ridge National Laboratory, Oak Ridge, Tennessee 37831, USA}

\author{Rebecca W. Smaha}
\affiliation{Stanford Institute for Materials and Energy Sciences, SLAC National Accelerator Laboratory, 2575 Sand Hill Road, Menlo Park, CA 94025, USA}
\affiliation{Department of Chemistry, Stanford University, Stanford, CA 94305, USA}

\author{Robert S. Feigelson}
\affiliation{Department of Materials Science and Engineering, Stanford University, Stanford, CA 94305, USA}

\author{Young S. Lee}
\affiliation{Department of Applied Physics, Stanford University, Stanford, CA 94305, USA}
\affiliation{Stanford Institute for Materials and Energy Sciences, SLAC National Accelerator Laboratory, 2575 Sand Hill Road, Menlo Park, CA 94025, USA}

\author{Rafael M. Fernandes}
\affiliation{School of Physics and Astronomy, University of Minnesota, Minneapolis,
MN 55455, USA}

\author{Ian R. Fisher}
\email{E-mail: irfisher@stanford.edu}
\affiliation{Department of Applied Physics, Stanford University, Stanford, CA 94305, USA}

\fi

\begin{abstract}
We report results of low-temperature heat capacity, magnetocaloric
effect and neutron diffraction measurements of TmVO$_{4}$, an insulator
that undergoes a continuous ferroquadrupolar phase transition
associated with local partially-filled $4f$ orbitals of the thulium
(Tm$^{3+}$) ions. The ferroquadrupolar transition, a realization of Ising
nematicity, can be tuned to a quantum critical point using a magnetic
field oriented along the $c$-axis of the tetragonal crystal lattice, which acts as an effective transverse field for the Ising-nematic order.
In small magnetic fields, the thermal phase transition can be well-described
using a semi-classical mean field treatment of the transverse field Ising model. However, in higher magnetic
fields, closer to the field-tuned quantum phase transition, subtle
deviations from this semi-classical behavior are observed due to quantum
fluctuations. Although the phase transition is driven by the local
$4f$ degrees of freedom, the crystal lattice still plays a crucial
role, both in terms of mediating the interactions between the local quadrupoles, and in determining
the critical scaling exponents, even though the phase transition itself
can be described via mean field. In particular, bilinear coupling
of the nematic order parameter to acoustic phonons changes the spatial and
temporal fluctuations of the former in a fundamental way,
resulting in different critical behavior of the nematic transverse-field Ising model
as compared to the usual case of the magnetic transverse-field Ising model. Our results establish TmVO$_{4}$
as a model material, and electronic nematicity as a paradigmatic example, for quantum criticality in insulators.
\end{abstract}

\ifpnas
\thispagestyle{firststyle}
\ifthenelse{\boolean{shortarticle}}{\ifthenelse{\boolean{singlecolumn}}{\abscontentformatted}{\abscontent}}{}
\else
\maketitle
\fi


\ifpnas
\dropcap{S}everal
\else
Several
\fi
experimental and theoretical studies indicate a
possible close association between nematic quantum criticality, non-Fermi-liquid
behavior and the occurrence of superconductivity (see, for example
references \citep{Shibauchi2014,Lederer2015,Metlitski2015,Lederer2017,Klein2018,Wang2018,Worasaran2021}
and references therein). This is, however, a complicated problem to
study, in large part because it is not clear how well the current
materials of interest map onto the effective models that are studied
theoretically. For instance, iron-based superconductors, which display an unambiguous nematic phase, usually display a closely related antiferromagnetic state as well \cite{Fernandes2014-01}. Furthermore, chemical substitution is often used as
the non-thermal control parameter that tunes candidate materials across
the putative nematic quantum critical point (QCP). This can be problematic,
since chemical composition is not a continuous variable, induces quenched
disorder, and can affect multiple terms in the effective Hamiltonian
in poorly understood and poorly controlled ways. All of these factors
motivate development of simpler model systems, for which key terms
in the underlying Hamiltonian are well understood and well-controlled,
and for which a nematic quantum phase transition (QPT) can be driven in
a continuous fashion in-situ without the need for chemical substitution.
Here, we consider the simplest case of a nematic QCP in an insulator, for which: 1) the nematic degrees of freedom are provided by local atomic orbitals, 2) the electron-hole excitations are gapped, and 3) the QCP can be traversed by application of a magnetic field.

Ferroquadrupolar (FQ) order of local $4f$ orbitals, in which each $4f$
orbital spontaneously acquires an electric quadrupole moment with
the same orientation below a critical temperature $T_{Q}$, is a specific
realization of electronic nematic order \citep{Maharaj2017,Rosenberg2019}.
Bilinear coupling between the local quadrupole moments and static
and dynamic lattice distortions with the same symmetry provides an
effective interaction between the local quadrupoles. Under
certain conditions, this interaction can then drive a cooperative phase transition to a
FQ ordered state. The FQ ordering is necessarily
accompanied by a structural distortion at the same critical temperature
$T_{Q}$ -- this is the essence of the cooperative Jahn-Teller effect
\citep{Gehring1975,Melcher1976}. For the case corresponding to a
tetragonal-to-orthorhombic phase transition, the FQ/nematic order parameter has
an Ising character and there are no cubic invariants in the free energy,
so the phase transition can be continuous. In the absence of disorder,
one anticipates mean field behavior for the thermal phase transition
since the upper critical dimension ($d_{c}^{+}$) for the Ising-nematic model
with strain-mediated long range interactions is 2 \citep{Karahasanovic2016,Paul2017}.
Under certain conditions, met for TmVO$_{4}$, a formal mapping to
the Transverse Field Ising Model (TFIM) can be made, in which a magnetic
field applied along the crystalline $c$-axis acts as an effective transverse
field for the local quadrupoles, effectively suppressing the quadrupole
order \citep{Maharaj2017}. 
What makes this system fundamentally different from the usual magnetic realization of the TFIM (as in, e.g. LiHoF$_4$ \cite{Bitko1996,Ronnow2005}) is the aforementioned bilinear coupling between the FQ order parameter and lattice deformations with the same symmetry. This coupling not only renders the thermal phase transition
mean field (by changing $d_{c}^{+}$ from 4 to 2, as noted above), but,
as we will show, also determines the critical behavior associated
with the field-tuned QPT \cite{Paul2017}.

\begin{figure}[!ht]
\includegraphics[width=1\columnwidth]{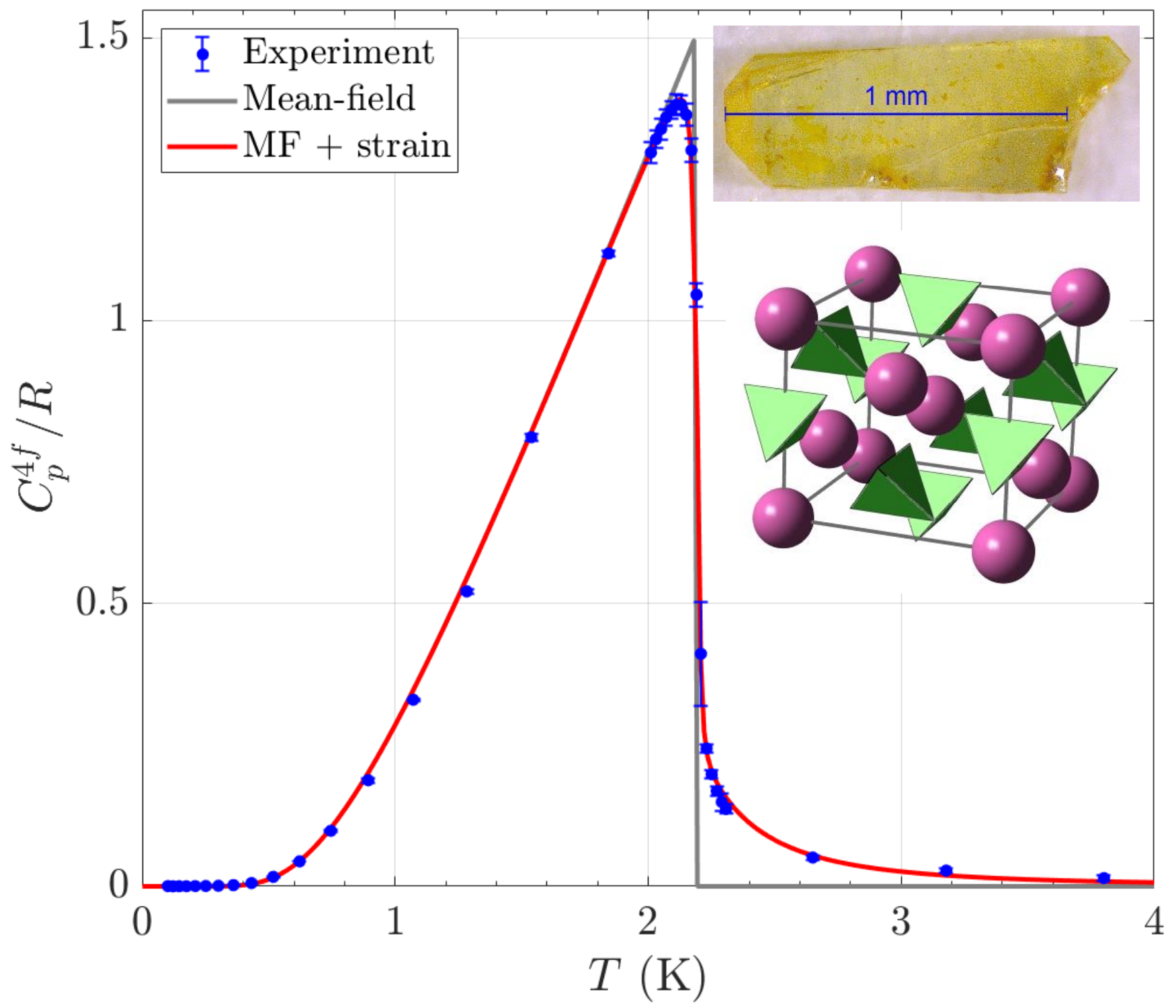}

\caption{Temperature dependence of the heat capacity of TmVO$_{4}$ in zero
magnetic field, illustrating the mean-field character of the phase transition. Data
(blue points) are shown after subtraction of the phonon thermal contribution,
and are normalized to the gas constant $R$. Gray line shows the mean-field
model, for which the only free parameter is the value of $T_{Q}$.
Red line includes phenomenological parameters to account for broadening
of the phase transition at $T_{Q}$ (modelled here by a small strain
of the same symmetry as the order parameter) and the contribution
to the heat capacity above $T_{Q}$ (modelled by a small temperature-independent
stress), as described in the Supplementary Material.
The upper inset shows an image of the sample on which these zero-field
data were measured. To minimize demagnetizing effects, needle-shaped
samples were used for measurements close to the field-tuned QCP. The
lower inset illustrates the zircon-type crystal structure in the tetragonal
phase, with Tm ions represented by purple spheres and VO$_{4}$ molecular
clusters by green coordination polyhedra.\label{fig:zero-field}}
\end{figure}

The title material, TmVO$_{4}$, is an insulator. It undergoes a continuous
FQ phase transition at $\unit[2.2]{K}$, with all the
action driven by the local partially filled $4f$ orbital of the Tm
ion. Material analogs that lack the partially filled $4f$ orbital,
such as YVO$_{4}$, do not undergo a similar phase transition, demonstrating
that the crystal lattice is perfectly stable in the absence of the
Jahn-Teller effect driven by local quadrupolar moments (see Supplementary Material).
At high temperatures, the material has tetragonal symmetry, with
the Tm ions occupying a unique crystallographic site with $\mathrm{D_{2d}}$
symmetry (inset of Figure \ref{fig:zero-field}). The ground state
of the Tm ion in the presence of the crystal electric field (CEF)
is a non-Kramers orbital doublet that transforms as the $E$ irreducible representation of the $\mathrm{D_{2d}}$ symmetry. It is this degeneracy
that drives the FQ phase transition. The first excited CEF state is $\unit[54]{cm^{-1}}$
($\unit[6.7]{meV}$) above the ground state doublet \citep{Knoll1971},
so at low temperatures the system can be described in terms of a pseudo-spin
to a very good approximation. Below $T_{Q}$, the material develops
spontaneous $\epsilon_{xy}$ strain, corresponding to a nematic order parameter that transforms as the $B_{2g}$ irreducible representation of the point group of the tetragonal crystal lattice. 
Thus, the principal axes of the resulting
orthorhombic state are rotated by 45 degrees with respect to those
of the high-temperature tetragonal structure.

This material was extensively studied in the 1970s and 1980s, in part
because of the “ideal” mapping to simple pseudo-spin models of the
cooperative Jahn-Teller effect. A series of beautiful measurements
established that the zero-field thermal phase transition of TmVO$_{4}$
is indeed mean-field-like, including measurements of heat capacity
(which shows the canonical mean-field “step” seen in Figure \ref{fig:zero-field})
\citep{Cooke1972}, optical absorption \citep{Becker1972}, Raman
spectroscopy \citep{Harley1972}, ultrasound \citep{Melcher1973,Melcher1974},
x-ray diffraction \citep{Segmueller1974}, M\"{o}ssbauer spectroscopy
\citep{Triplett1974}, inelastic neutron scattering \citep{Kjems1975},
electron paramagnetic resonance \citep{Schwab1975}, and nuclear magnetic
resonance \citep{Bleaney1980,Wang2021TmVO4}. Indeed, the material was seen as a
model system for mean-field thermal phase transitions \citep{Melcher1976}.
Similarly, the shape of the phase boundary in the $H-T$ plane appeared
to conform to mean-field expectations in the temperature range considered,
based on a simple semi-classical solution of the TFIM. In this approach, the quantum dynamics
associated with the non-commuting pseudo-spin operators in the TFIM are neglected, such that the external transverse field and the intrinsic Weiss longitudinal field experienced by the local quadrupoles are simply added
in quadrature \citep{Stinchcombe1973a}. However, when the material
was first investigated, notions of both quantum phase transitions
and electronic nematic order had yet to be considered in detail in
the context of condensed matter. As a result, the quadrupole phase transition
and associated physical properties were not followed to low temperatures,
proximate to what we would now understand to be a putative FQ
QCP. Recent theoretical studies of electronic nematic
order, inspired by materials such as the Fe-based superconductors,
have underscored the importance of the bilinear coupling of order parameter
fluctuations to lattice deformations with the same symmetry \citep{Paul2011,Zacharias2015,Paul2017}.
Indeed, as we will show, this coupling profoundly affects the resulting critical behavior that characterizes the continuous mean field QPT.%


In this article, we report results of measurements of TmVO$_{4}$
that probe the field-dependence of the order parameter at low temperatures,
and the shape of the phase boundary approaching the field-tuned QPT.
Far from the QPT, the material is well-described by the semi-classical
treatment of the TFIM discussed above, in which dynamics associated
with the non-commuting operators is neglected. Closer to the QPT,
however, subtle deviations are evident, arising as a consequence
of the quantum fluctuations. We discuss expectations for the associated
mean-field QPT. In particular, because the dimensionality
$d$ is greater than $d_{c}^{+}$ for both the classical and quantum
phase transitions, we can obtain $T_{Q}$ as a function of $\left(H-H_{c}\right)$
by computing the Gaussian-fluctuations corrections to the free energy, which are well
controlled in this case. Bilinear coupling of the FQ order parameter
to phonon modes with the same symmetry necessarily changes
these exponents relative to the case of the magnetic TFIM. Our experimental
observations and associated theoretical treatment elevate TmVO$_{4}$
to the status of a model material system, not only to reveal mean-field behavior at the thermal phase transition in zero field (the
primary result from the 1970’s), but now also to elucidate the effects
of quantum critical fluctuations proximate to the mean field continuous
FQ QPT.

\section{Results}

\subsection{Far from the QCP}

The cooperative Jahn-Teller effect in TmVO$_{4}$ results in a pseudo-proper
ferroelastic phase transition. Bilinear coupling between the FQ order parameter and the $\epsilon_{xy}$ orthorhombic lattice distortion
ensures that the temperature and field-dependence of the lattice distortion,
measured here via elastic neutron scattering, is the same as that
of the local quadrupole moments. The orthorhombicity, defined as $\delta=\left|a-b\right|/\left(\left(a+b\right)/2\right)$,
was determined from the splitting of the $\left(880\right)_{T}$ Bragg
peak, where the subscript $T$ denotes $\left(hkl\right)$ labelling
according to the \emph{high-temperature tetragonal} unit cell, whereas
$a$ and $b$ are the lattice parameters in the \emph{low-temperature
orthorhombic} phase. Measurements were performed at a temperature $T = \unit[0.60]{K} \approx 0.27\, T_{Q}$ as a function of magnetic
field $H$, with the field oriented along the crystalline $c$-axis.
Demagnetization effects were modelled using a finite elements simulation,
discussed in greater detail in the Supplementary Material.

\begin{figure}[!ht]
\includegraphics[width=1\columnwidth]{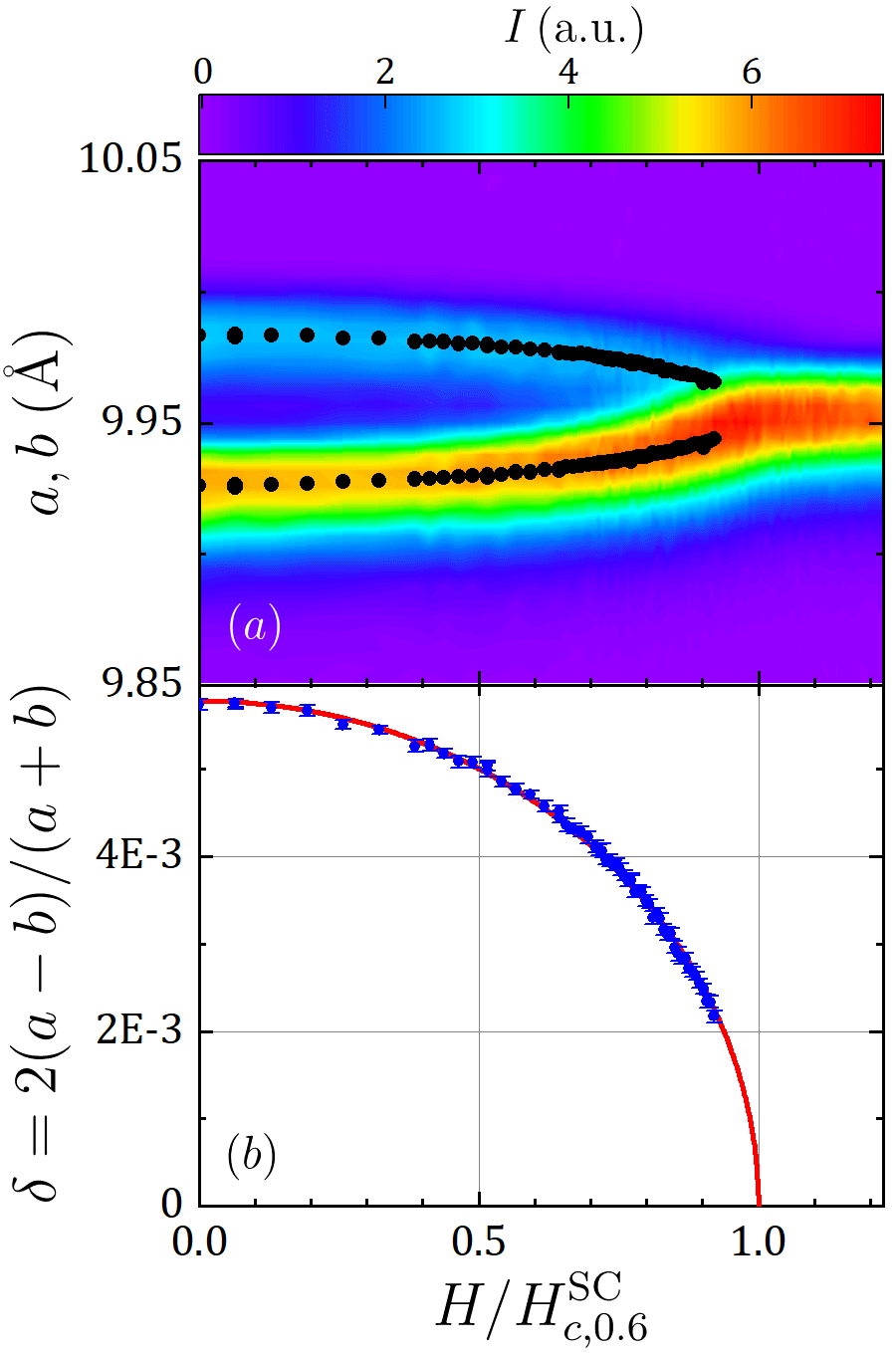}

\caption{Field-dependence of (a) the lattice parameters $a$ and
$b$ in the orthorhombic phase, and (b) the orthorhombicity parameter $\delta$ at $T=\unit[0.60]{K}$.
$H_{c,0.6}^{\mathrm{SC}}$ is the critical field derived from the semi-classical fit described in equation
\ifpnas
[\ref{eq:order_parameter}]
\else
(\ref{eq:order_parameter})
\fi
\negmedspace
\negmedspace
.
The color scale in panel (a) corresponds to detector intensities integrated
over the transverse scattering direction. Superimposed are the data
points indicating the positions of the peak maxima (black dots). Differences
in the intensity of the two peaks are due to differences in the domain
population. The orthorhombicity parameter closely follows the semi-classical mean-field
solution for the order parameter of the TFIM (red
line in panel (b)).\label{fig:orthorhombic_distortion}}
\end{figure}

As shown in Figure \ref{fig:orthorhombic_distortion}(a), increasing
$H$ results in a continuous suppression of the order parameter for
the ferroelastic distortion, with the two Bragg peaks, associated
with the two orthorhombic domains, eventually merging into a single
peak. Peaks for each value of magnetic field were fitted using a
standard neutron time-of-flight functional form \citep{Ikeda1985,Rodriguez-Carvajal1993},
in order to identify the positions $x_{M,1}$ and $x_{M,2}$ of their
maxima, from which the relative distortion was deduced as $2\left|x_{M,1}-x_{M,2}\right|/\left|x_{M,1}+x_{M,2}\right|$
(see Supplementary Material).
Its dependence on the magnetic field $H$, plotted in Figure \ref{fig:orthorhombic_distortion}(b),
closely follows what is expected from semi-classical mean-field treatments of the TFIM, where the transverse (magnetic) and longitudinal ($B_{2g}$ strain) fields add in quadrature \citep{Stinchcombe1973a}:

\begin{equation}
\delta(T=\unit[0.6]{K},H) = \delta_{0}(\unit[0.6]{K})\cdot\sqrt{1-\left(\frac{H}{H_{c,0.6}^{\mathrm{SC}}}\right)^{2}}\label{eq:order_parameter}
\end{equation}
where $\delta_{0}(\unit[0.6]{K})$ is the orthorhombic distortion
in the absence of magnetic field, and $H_{c,0.6}^{\mathrm{SC}}$
is the critical field 
at $T=\unit[0.6]{K}$, derived from this semi-classical fit.

Equation
\ifpnas
[\ref{eq:order_parameter}]
\else
(\ref{eq:order_parameter})
\fi
fits the data very well up to
$H/H_{c,0.6}^{\mathrm{SC}} \approx 0.92$, with no systematic deviations
between the data and the semi-classical mean-field solution discernible
within experimental error. For fields above this value, the peak-fitting
is not constrained enough to give reliable values due to the instrumental
resolution. The semi-classical mean-field fit yields $\delta_{0}(\unit[0.6]{K}) = 5.78(4)\cdot10^{-3}$,
in agreement with previous estimates for the zero-temperature orthorhombicity based on the same semi-classical model
\citep{Segmueller1974}.

A further test of the extent to which TmVO$_{4}$ follows semi-classical
mean-field expectations as a function of field is provided by the
shape of the phase boundary, which is best determined by thermodynamic
probes. Heat capacity measurements were performed in a Helium-3 refrigerator
down to $\unit[0.35]{K}$.

\begin{figure}[!ht]
\includegraphics[width=1\columnwidth]{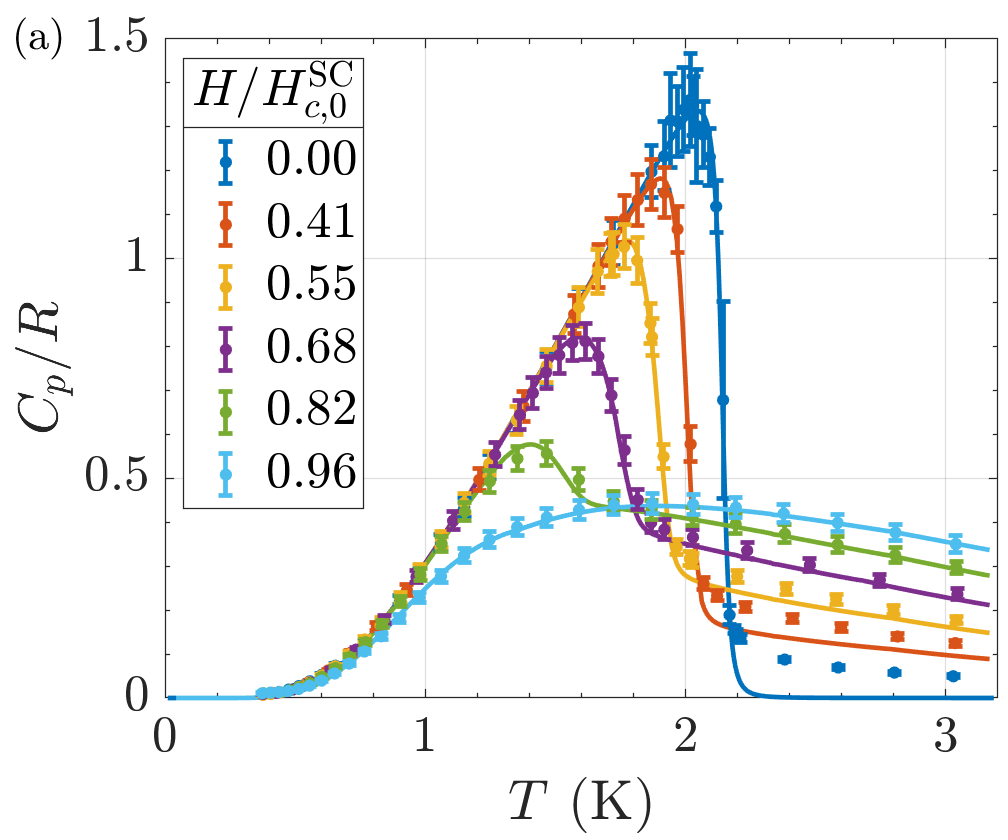}

\vspace{2mm}

\includegraphics[width=1\columnwidth]{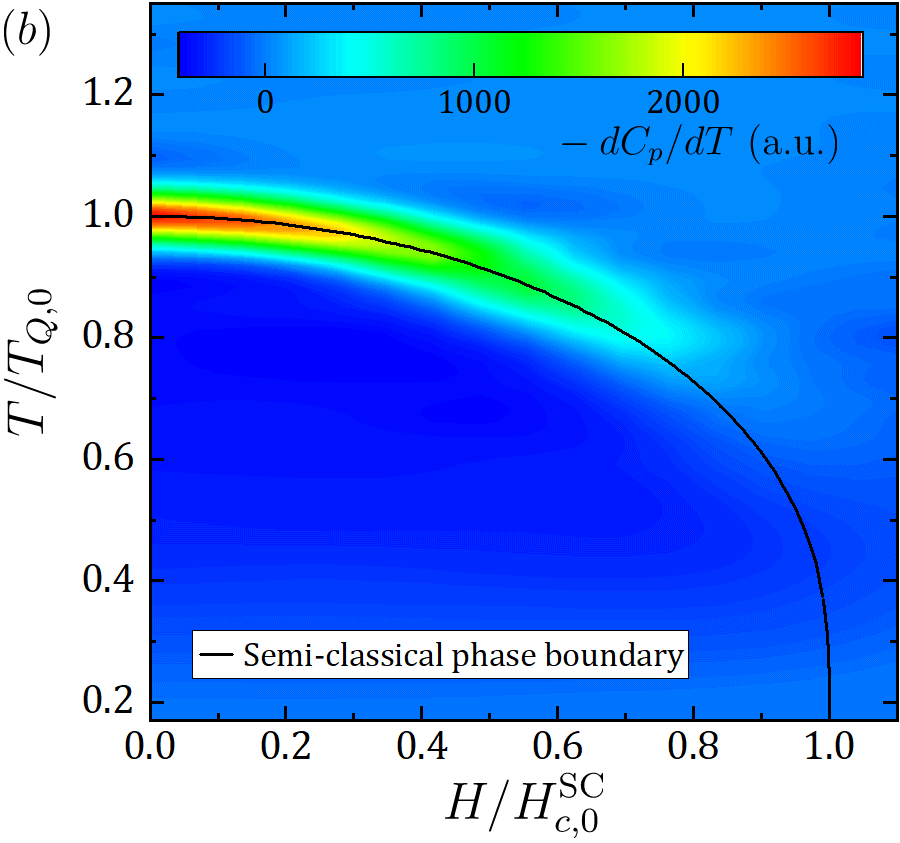}

\caption{(a) Temperature dependence of the heat capacity of TmVO$_{4}$ under various
values of externally applied magnetic field $H$, in units of the
gas constant $R$. Symbols: experimental data. Full line at zero field:
mean-field fit using the semi-classical TFIM. All other solid lines:
mean-field computations of the semi-classical solution of the TFIM, combined with the distributions
of magnetic fields inside the sample, as computed using COMSOL (see
text). (b) Temperature--magnetic field phase diagram as obtained
from the first derivative of the heat capacity (color map). The black
line is the theoretical phase boundary according to the semi-classical mean-field solution of the
TFIM. $T_{Q,0}$ is the transition temperature in the absence of magnetic
field; $H_{c,0}^{\mathrm{SC}}$ is the critical transverse magnetic field at
$T=0$, as extracted from fitting the heat capacity data with the
semi-classical expression for the phase boundary.\label{fig:Cp}}
\end{figure}

The temperature and field dependence of the heat capacity (data points
in Figure \ref{fig:Cp}(a)) closely agree with the semi-classical mean-field solution of the TFIM \citep{Stinchcombe1973a,Stinchcombe1973c}.
In particular, they exhibit three main features as the magnetic field
is increased: 1) The step marking the phase transition moves to lower
temperatures, and is suppressed in magnitude; 2) The data fall onto
the same curve below the transition temperature $T_{Q}(H)$; and 3)
The high-temperature tail of a Schottky anomaly is clearly visible
above the transition temperature due to splitting of the ground state
doublet induced by the magnetic field.

Rounding of the mean-field step in the heat capacity is evident in
Figures \ref{fig:zero-field} and \ref{fig:Cp}(a). This effect is
attributed to small unintentional strain of the same symmetry as the order parameter and, for measurements made in magnetic fields,
demagnetization effects. In zero field, the upward curvature of the
heat capacity for temperatures \emph{above $T_{Q}$} can be
accounted for by a small residual stress, which could arise, for example,
from freezing of the grease used to hold the sample to the calorimeter
platform, or possibly even from crystal growth defects.
While both
sources of strain are likely inhomogeneous, the data can be very well described
by treating both by a single uniform parameter (solid red line in
Figure \ref{fig:zero-field}). These subtle effects are of course
still present when the measurements are made in an applied field,
but are swamped by the much larger field-induced effects associated
with inhomogeneous demagnetization and field-induced splitting above
$T_{Q}$.

Demagnetizing effects were modelled using finite element simulations
(see Supplementary Material).
This results in a very good description of the data through $T_{Q}$
for all fields (solid lines in Figure \ref{fig:Cp}(a))
\footnote{The constant stress underlying the phenomenological fit of the high-temperature tail of heat capacity, as shown in Figure \ref{fig:zero-field}, is independent of magnetic field. 
Since we focus here on the effects of magnetic field on the heat capacity, we do not include this high-temperature contribution in our modelling.}.
It is worth noting that only the solid (zero field) blue line is an actual fit of the corresponding data, with two free parameters, $T_{Q,0}$, the zero-field transition temperature, and $\varepsilon$, the \emph{homogeneous} longitudinal field responsible for the rounding of the transition.
All other solid lines at finite fields
were computed numerically \emph{without any free parameters} using: 1) the values obtained for $T_{Q,0}$ and $\varepsilon$ from the fit of the zero-field data, and 2) the magnetic
field distributions computed in COMSOL.

For magnetic fields up to $\sim0.9\, H_{c}(T=0)$, the
critical temperature can be readily extracted from the first temperature
derivative of the heat capacity data, $dC_{p}/dT$, which is plotted
as a color map on the $T-H$ plane in Figure \ref{fig:Cp}(b). The same figure also displays 
the semi-classical mean-field phase boundary (black line), which is described
by the following functional form \citep{Stinchcombe1973a}:

\begin{equation}
\frac{T_{Q}\left(H\right)}{T_{Q,0}}=\frac{H/H_{c,0}^{\mathrm{SC}}}{\atanh\left(H/H_{c,0}^{\mathrm{SC}}\right)}\label{eq:semi-classical}
\end{equation}
where $T_{Q,0}$ is the zero-field transition temperature, and $H_{c,0}^{\mathrm{SC}}$
is the zero-temperature critical field. This equation fits the phase boundary obtained from heat capacity
data very well for fields up to $\sim0.8\, H_{c,0}^{\mathrm{SC}}$.

\subsection{Close to the QCP}

The small magnitude of the heat capacity anomaly for higher fields,
combined with the steep shape of the mean-field phase boundary, render
heat capacity measurements less helpful for determining $T_{Q}$ closer
to the QPT. For this reason, we instead use the Magnetocaloric
Effect (MCE).

MCE measurements were made using the same calorimeter as the heat
capacity measurements in the He-3 fridge, and also using a different
calorimeter in a separate dilution refrigerator. Representative data,
along with a description of the physical principle of the MCE, can
be found in the Supplementary Material.
Under quasi-adiabatic conditions, to compensate for the release of entropy associated
with the quadrupolar degrees of freedom at the FQ transition, the temperature
of the sample changes. For each MCE trace,
$H_{c}(T)$ was thus identified as the magnetic field value at which
the second derivative of the field-induced temperature change of the sample $\frac{d^{2}\Delta T}{dH^{2}}$ is maximum.%

\begin{figure}[!ht]
\includegraphics[width=1\columnwidth]{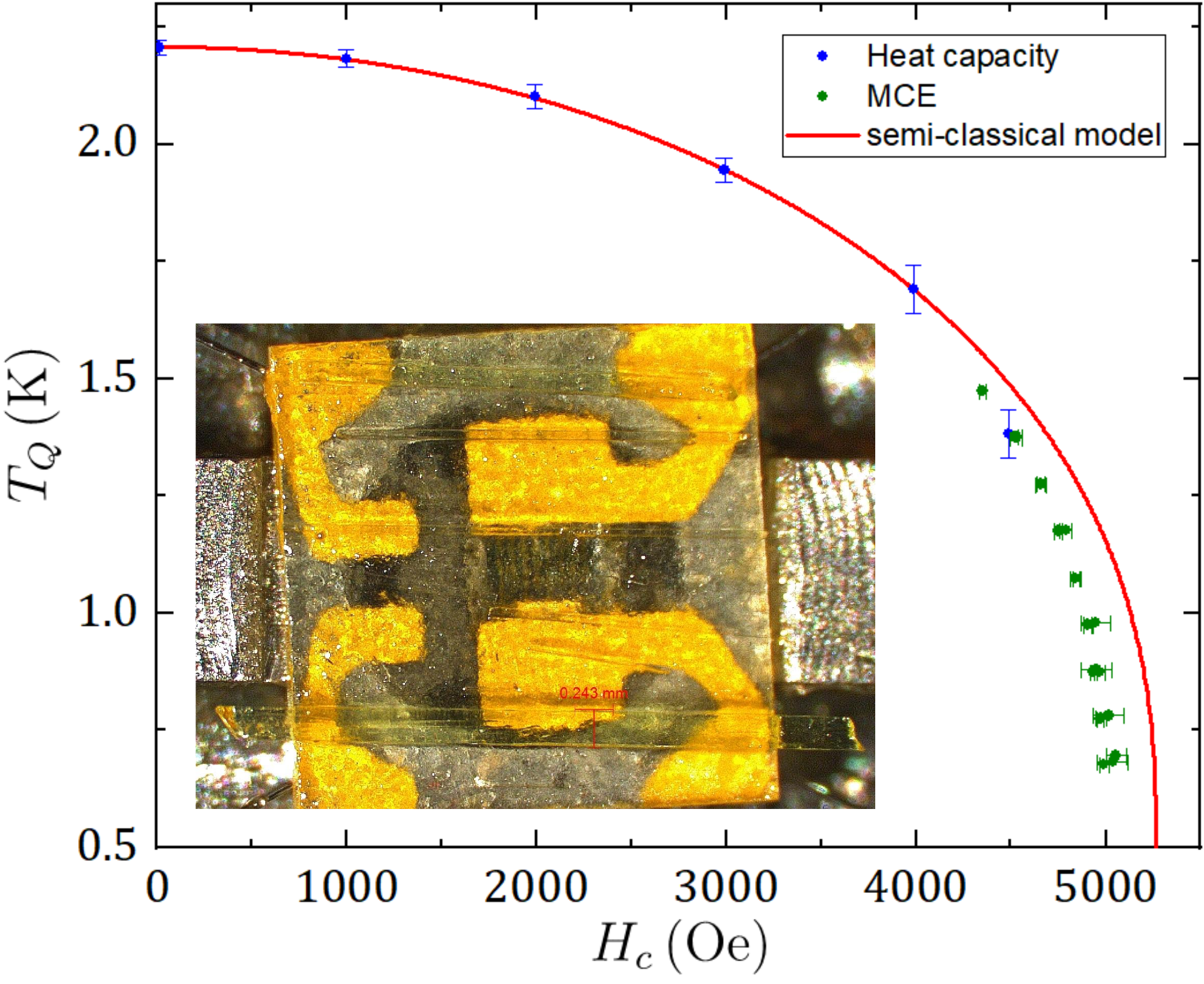}

\caption{Experimentally determined phase boundary $T_{Q}(H)$ extracted from measurements
of heat capacity (blue data points with vertical error bars) and magneto-caloric
effect (green data points with horizontal error bars). Red line shows
the fit of the phase boundary using the semi-classical solution of the TFIM, including data points at
or below $\unit[4]{kOe}$ only (see main text). The inset shows a
picture of the sample, made of a collection of needles to minimize
demagnetization effects, on the heat capacity platform.\label{fig:phase_boundary}}
\end{figure}

The values of $H_{c}(T)$ extracted via MCE measurements are plotted in Figure
\ref{fig:phase_boundary} as green dots with horizontal error bars,
along with the transition temperatures $T_{Q}(H)$ extracted
from the heat capacity data, plotted as blue
dots with vertical error bars. The combination of all these data into
a single dataset was then fitted on various ranges of magnetic field,
using the expression for the phase boundary in the semi-classical description of the
TFIM, Eq.
\ifpnas
[\ref{eq:semi-classical}]
\else
(\ref{eq:semi-classical})
\fi
\negmedspace
\negmedspace
. The best fit was obtained
when fitting over the range $0\leq H\leq\unit[4]{kOe}$, with an adjusted-$R^{2}$
of 0.9998. The corresponding curve is plotted as a red solid line
in figure \ref{fig:phase_boundary}. While in the field range between 0 and $\unit[4]{kOe}$,
the curve goes through all the data points, it overshoots the data for fields larger than $\unit[4]{kOe}$, when the QCP is approached.

It is worth noting that this result is not an artefact of data obtained
from two different types of measurements, as the heat capacity data
point located at $H=\unit[4.5]{kOe}$ is in agreement with the MCE
data. Moreover, both measurements were performed in a single run using
the same setup. Lastly, these results were reproduced on the sample
of Figure \ref{fig:zero-field}.

\begin{figure}[!ht]
\includegraphics[width=1\columnwidth]{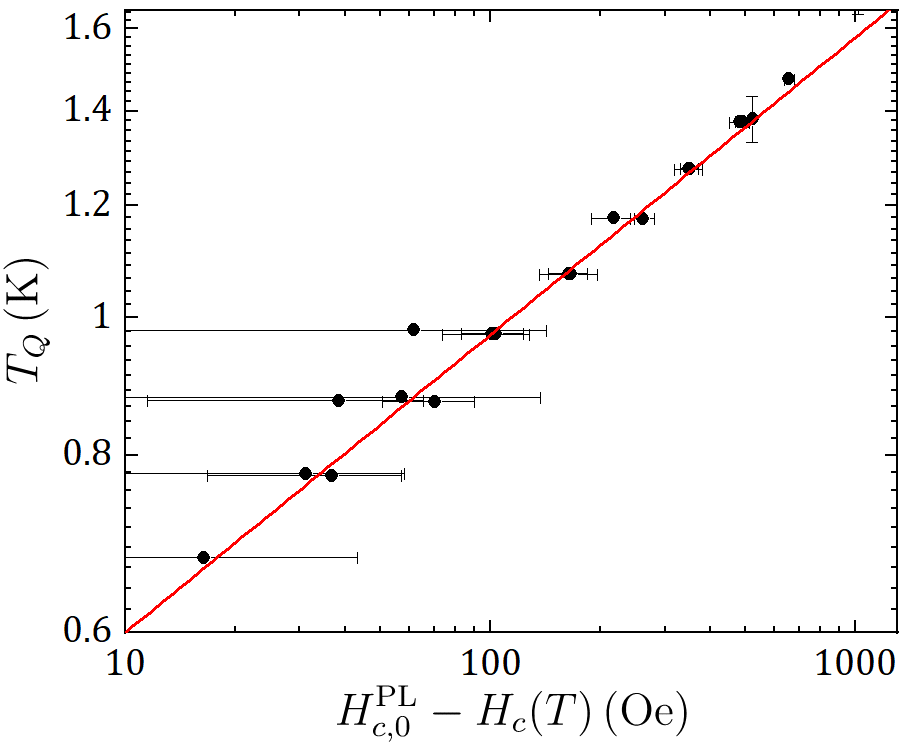}

\caption{Log-log plot of the phase boundary $T_{Q}(H)$.
The experimental
data (black dots with error bars) were fitted by a power law, $T_{Q}\sim\left(H_{c,0}^{\mathrm{PL}}-H_{c}(T)\right)^{\psi}$.
The best fit, plotted as a red solid line, yields $\psi=0.21(2)$ and $H_{c,0}^{\mathrm{PL}}=\unit[5008(4)]{Oe}$.
The latter value is used as reference for the horizontal axis of the figure.\label{fig:Log-log-plot}}
\end{figure}

Since the data deviate from the semi-classical mean-field solution of the TFIM, we used a self-consistent Gaussian-fluctuation approach to calculate the expected shape of the phase boundary.
This method includes the leading order contribution of the quantum fluctuations to $T_{Q}(H)$, which should be sufficient since the system is above the upper critical dimension \cite{Loehneysen2007}. Our model predicts a power-law
behavior close to the QCP (see below and the Supplementary Material),
i.e. $T_{Q}\sim\left(H_{c,0}^{\mathrm{PL}}-H_{c}(T)\right)^{\psi}$, where $H_{c,0}^{\mathrm{PL}}$ denotes the estimated critical field at zero temperature as determined from the power law fit.
Fitting the data of figure \ref{fig:phase_boundary} with this expression
yields $H_{c,0}^{\mathrm{PL}}=\unit[5008(4)]{Oe}$ and $\psi=0.21(2)$,
with a best fit range $\unit[4.5]{kOe}\le H\le\unit[5.0]{kOe}$ and
an adjusted-$R^{2}$ of 0.97. The resulting best fit curve is shown
as a red solid line on top of the data (black dots) in the log-log
plot of Figure \ref{fig:Log-log-plot}.

We note that the neutron scattering data shown in Figure \ref{fig:orthorhombic_distortion} are only able to resolve the orthorhombicity over a regime of field and temperature that predominantly lies outside (and barely overlaps with) the proposed quantum critical regime where power law behavior of the critical temperature is observed.
This is a consequence of the very steep phase diagram. 
Nevertheless, if the orthorhombicity could be resolved closer to $H_c$, this too, must deviate from the semi-classical model in order to match the observed phase diagram.

\section{Discussion}

Power law behavior of the phase boundary $T_{Q}\sim\left(H_{c,0}^{\mathrm{PL}}-H_{c}(T)\right)^{\psi}$
is a characteristic feature of quantum criticality \citep{Loehneysen2007}.
The associated quantum fluctuations, presumably promoted by the applied
magnetic field acting as a transverse nematic field \citep{Maharaj2017},
suppress the critical temperature below what would be anticipated
based solely on the semi-classical solution described above. Significantly,
bilinear coupling of the Ising-nematic order parameter to dynamic
$\epsilon_{xy}$ shear deformations makes this fundamentally
different from the magnetic TFIM case, as widely studied for instance
in LiHoF$_{4}$ \citep{Bitko1996,Ronnow2005}. Indeed, as we show below, the
QCP is characterized by different scaling exponents.

It is well-established that the thermal phase transition for Ising-nematic order is mean-field like due to the long-range nematic interaction
generated by the coupling to acoustic phonons \citep{Qi2009,Karahasanovic2016,Paul2017}.
Because the correlation length only diverges along specific in-plane directions in momentum space
(at 45 degrees with respect to the principal axes of the nematic distortion),
the upper critical dimension of the problem is reduced from 4 (for the simple Ising
model) to 2. Equivalently, the effective dimensionality $d$ of
the system increases from 3 to 5 due to this coupling, i.e. the material
behaves like an Ising system in $d=5$. Proximate to the QCP, the
effective dimensionality increases yet again to
$d+z$ \citep{Loehneysen2007}, where $z$ is the dynamical critical
exponent that characterizes temporal fluctuations of the order parameter
near the QCP. In other words, since the thermal phase transition for
Ising-nematic order in a compressible lattice is mean-field like,
the QCP should also be.

Since mean-field behavior is observed across the entire phase diagram
and the Ginzburg criterion is never violated, a one-loop self-consistent
approximation is sufficient to describe how the transition temperature
is suppressed to zero by Gaussian quantum fluctuations in an insulator
\citep{Loehneysen2007}. For strong nemato-elastic coupling, appropriate
for this system in which the nematic order only exists because of
coupling to the lattice, this calculation results in an exponent that is consistent
with the standard scaling relation $\psi=z/(z+d-2)$ \citep{Loehneysen2007}
with $d=5$ (see Supplementary Material). However, since the phase boundary is measured over a range
of temperatures and magnetic fields, different exponents can be manifested
in different temperature/magnetic-field regimes, corresponding to
different values of $z$, while still belonging to the same mean-field
description. To understand why $z$ changes as the QCP is approached,
we note that in an insulator the critical dynamics of a generic bosonic
mode is expected to be propagating, resulting in $z=1$. However,
a quantum elastic transition has an exponent $z=2$ \citep{Folk1979,Zacharias2015}.
This follows from the fact that the dispersion relation $\omega(q)$
for the soft acoustic phonons (with wave-vector along $[100]$ and
$[010]$) at the critical field is only linear for high $q$ values,
crossing over to quadratic behavior approaching $q=0$ due to the vanishing
of the sound velocity \citep{Weber2018,Merritt2020}. Consequently,
as temperature is decreased and the nematic QCP is approached, a crossover
is anticipated from a regime in which temporal fluctuations are characterized
by $z=1$ to one characterized by $z=2$. The associated exponent
describing the phase boundary in these two regimes is then expected
to change from $\psi=1/4$ ($d=5$, $z=1$) at higher temperatures to $\psi=2/5$
($d=5$, $z=2$) at lower temperatures. The crossover scale is set
not only by the nemato-elastic coupling, which also affects the crossover
from $d=3$ to $d=5$, but also by the ratio between the velocities
of propagation of sound and of the nematic mode (see Supplementary Material). As a result, the
crossover scale from $d=3$ to $d=5$, associated with spatial fluctuations,
does not need to coincide with the crossover scale from $z=1$ to
$z=2$ associated with temporal fluctuations.

The observed exponent $\psi=0.21\pm0.02$ is remarkably close to $1/4$,
the value anticipated for $d=5$ and $z=1$. Under the assumption
that this apparent power law behavior originates from critical scaling,
we deduce that for the temperature range over which the measurements
were performed, $z=1$. This would be consistent with a small ratio
between the bare nematic mode velocity and the bare sound velocity,
i.e. the purely electronic nematic fluctuations propagate slower than
sound does. The fact that the local $4f$ quadrupoles are essentially only coupled via the lattice supports this assessment. Presumably, extending our measurements to progressively
lower temperatures would reveal a crossover to a regime in which the
$z=2$ exponent is eventually observed.

Our result demonstrates the significance of nemato-elastic coupling
for Ising-nematic systems. For the simpler case of the \emph{magnetic} TFIM,
the anticipated exponent describing the phase boundary close to the
QCP, obtained from the same self-consistent calculation, is $\psi=1/2$ ($d=3$, $z=1$). In that case, the Ising-magnetic order parameter couples only quadratically
to longitudinal strain, which is predicted to result in a first-order transition
\citep{Larkin1969,Bergman1976}. In the Ising-\emph{nematic} case however, the effects of a compressible
lattice are fundamentally different:
here, the bilinear coupling of the nematic order parameter to shear strain
means that the nematic mode inherits some key properties of the critical elasticity of the lattice \citep{Zacharias2015},
such as a strong anisotropy of the correlation length, although the
phase transition itself remains continuous \citep{Karahasanovic2016,Paul2017}.
Our observation of the associated power law proximate to the QCP underscores
a key point that applies to metallic nematic systems just as much
as for insulators: sufficiently close to the QCP (where ‘sufficiently
close’ could depend on microscopic details for different material
systems), coupling to the lattice fundamentally affects the fluctuating
order and cannot be neglected \citep{Zacharias2015,Paul2017,Carvalho2019}.
More broadly, our results establish the Ising-nematic QPT in insulators such as TmVO$_4$ as a paradigmatic framework to elucidate quantum criticality, which is complementary to, but qualitatively different from, the standard example of an Ising ferromagnet subjected to a transverse magnetic field \citep{Sachdev2011}.%



\ifpnas
\matmethods{
\else
\subsection*{Materials and Methods}
\fi
For heat capacity and magneto-caloric effect, single crystals of TmVO$_{4}$
were grown in a flux of Pb$_{2}$V$_{2}$O$_{7}$ using 4 mole percent
of Tm$_{2}$O$_{3}$, following the methods described in \citep{Feigelson1968,Smith1974}.
Samples were characterized using heat capacity measurements and show
high-quality with a sharp transition at $\unit[2.15\left(5\right)]{K}$
under zero magnetic field (see Supplementary Material).
For neutron scattering measurements, which require bigger single crystals, the
samples were grown using the floating zone method, as reported in
\citep{Oka2006}.
The crystal structure was verified on a flux-grown sample, by collecting single crystal X-ray diffraction data at beamline 12.2.1 at the Advanced Light Source, Lawrence Berkeley National Laboratory, which confirmed that the tetragonal $I4_1/amd$ symmetry observed at room temperature \citep{Chakoumakos1994} persists down to at least \unit[100]{K} (See Supplementary Material).
A Crystallographic Information File (CIF) has been deposited in the Cambridge Crystallographic Data Center (CCDC) with accession code 2117139.

Heat capacity and MCE measurements were performed in vacuum using the same setup, with magnetic field parallel to the $c$-axis of the sample, and without cycling temperature between one type of measurement and the other.
Heat capacity measurements were carried out under constant magnetic field, from $\unit[4]{K}$ down to base temperature ($\unit[0.1]{K}$ for the dilution refrigerator, $\unit[0.35]{K}$ for the Helium-3 cryostat).
For MCE measurements, bath temperature was constant, and magnetic field was swept at a constant rate
from 0 to $\unit[10]{kOe}$
(see Supplementary Material).

Elastic neutron scattering experiments were carried out in a dilution
refrigerator at the CORELLI beamline of the Spallation Neutron Source
at Oak Ridge National Laboratory. 
The sample was aligned in the $\left(HK0\right)$ scattering plane, 
with the magnetic field applied vertically along the $c$ axis.
Data were fitted using a convolution of a pseudo-Voigt function with the Ikeda-Carpenter function, which
is common for time of flight neutron scattering experiments \citep{Ikeda1985}
(see Supplementary Material).

Modeling of the sample measured under magnetic field was made using
the AC/DC module of the COMSOL$^{\text{®}}$ software \citep{Comsol55}.
\ifpnas
}

\showmatmethods{} 

\acknow{
\else
\section*{Acknowledgements}
\fi

Heat capacity and magnetocaloric effect measurements performed at Stanford University were supported by the Air Force Office of Scientific Research under award number FA9550-20-1-0252. 
Crystal growth experiments were supported by the Gordon and Betty Moore Foundation Emergent Phenomena in Quantum Systems Initiative through Grant GBMF9068. The neutron scattering activities were supported by the U.S. Department of Energy (DOE), Office of Science, Basic Energy Sciences, Materials Sciences and Engineering Division, under contract DE-AC02-76SF00515. 
A portion of this research used resources at the Spallation Neutron Source (SNS), a DOE Office of Science User Facility operated by the Oak Ridge National Laboratory (ORNL), and resources of the SNS Second Target Station Project. ORNL is managed by UT-Battelle LLC for DOE’s Office of Science, the single largest supporter of basic research in the physical sciences in the United States.
A portion of this research used resources of the Advanced Light Source, a U.S. DOE Office of Science User Facility under contract no. DE-AC02-05CH11231.
R.W.S. was supported by a National Science Foundation Graduate Research Fellowship (DGE-1656518).
Theory work (R.M.F.) was supported by the U. S. Department of Energy, Office of Science, Basic Energy Sciences, Materials Sciences and Engineering Division, under Award No. DE-SC0020045.
We acknowledge helpful discussions with Yuval Gannot, Marcus Garst, Steve Kivelson and Indranil Paul.
This manuscript has been authored by UT-Battelle, LLC, under contract DE-AC05-00OR22725 with the US Department of Energy (DOE). The US government retains and the publisher, by accepting the article for publication, acknowledges that the US government retains a nonexclusive, paid-up, irrevocable, worldwide license to publish or reproduce the published form of this manuscript, or allow others to do so, for US government purposes. DOE will provide public access to these results of federally sponsored research in accordance with the DOE Public Access Plan (http://energy.gov/downloads/doe-public-access-plan).
\ifpnas
}

\showacknow{} 
\fi

\bibliography{./Bibliography/biblio_CJTE,./Bibliography/Techniques,./Bibliography/biblio_FeSC,./Bibliography/General_Cond_Mat}

\end{document}